# ON SUB-GRID SCALE MODELING IN A DIFFERENTIALLY HEATED CUBICAL CAVITY USING COARSE LARGE EDDY SIMULATION

M.A. Sayed[1,2], A. Dehbi[1], M. Hadžiabić[3], B. Ničeno[1] and K. Mikityuk[1,2]

[1]*Paul Scherrer Institut, CH-5232 Villigen PSI, Switzerland*
[2]*Swiss Federal Institute of Technology Lausanne (EPFL), 1015 Lausanne, Switzerland*
[3]*International University of Sarajevo, Sarajevo 71000, Bosnia & Herzegovina*

Mohamed.sayed@psi.ch

**Abstract**

Two widely used sub-grid scale models, the standard and the dynamic Smagorinsky models were tested in a simulation of the flow in a thermally driven 3D cavity at Rayleigh number of Ra=$10^9$. The main focus of the research is the response of the subgrid scale models to the coarse mesh and their ability to predict the flow in a thermally driven 3D cavity with the mesh resolution below generally accepted LES standards. The research is motivated by investigating a feasible modeling strategy for the particulate flow in the wall-bounded side-wall heated cavity. As URANS and hybrid RANS-LES models fail to produce modeled energy due to laminarization, the alternative is LES applied on the coarse mesh. In a quantitative manner, first and second-moment statistics are compared against both reference LES and experimental databases from literature. For first moment statistics, both models globally predict the flow field well. However, for higher moments, the dynamic model outperforms the standard Smagorinsky model using the same mesh resolution. In a qualitative fashion, we report the contours of both temperature and velocity fields as well as the near-wall turbulent coherent structures. Results of flow quantities are globally in very good agreement with both reference LES and experimental measurements at a fraction of the CPU power needed for conventional LES.

## 1 Introduction

Particle transport in containments is very crucial for many engineering, medical and environmental applications from flow inside hospital rooms and the transport of air-borne pollutants in clean rooms to radioactive particulate flows in nuclear plants and aircraft-related applications. [Morrison, G. C. et al, 2006; Kam, T. et al., 1998; Wana, M. P. et al., 2007; Poussou, S. B. et al., 2010; Jones, A. V. et al., 2000]. For that sake, having a reliable design-based tool to predict such flows is very important to reduce the cost relative to the prototype-based approach. A branch of particulate flows in closed containments is thermally-driven flows in a parallelepiped cavity due to buoyancy circulations. Such configuration is referred to as a Differentially Heated Cavity (DHC), where two opposite walls are put to two different temperatures and the rest of the walls are adiabatically insulated.

Over the last few decades, numerous studies have investigated the flow of DHC both numerically and experimentally. In particular, the first numerical simulation of flow inside a DHC date back to 1983, when De Vahl Davis [G. De Vahl Davis, 1968] conducted a study on 2D square cavities with laminar flow. With the exponential development in computer power, research has found its way to endeavor 2D and 3D turbulent flows inside DHC with moderate to very high Rayleigh numbers with both Large Eddy Simulation (LES) and Direct Numerical Simulation (DNS); some of which are [A. Dehbi et al, 2017; Puragliesi R., 2010; Sebilleau F. et al., 2018; Sergent A., 2013]. Currently, the DHC serves as one of the main benchmarks used to validate turbulent buoyancy-driven flows in CFD. In this paper, we solely focus on the gas flow prediction in an extensive analysis as a first step prior to investigating the discrete phase.

To properly predict particle dispersion, the underlying carrier flow must be well-predicted. Despite being a simple geometry, DHC poses a big challenge for CFD codes to accurately capture the particulate flow behavior in a 3D turbulent regime. The main challenge for both the Unsteady Reynolds-Averaged Navier-Stokes (URANS) and hybrid RANS-LES models is the simultaneous existence of both laminar and turbulent regions and the underlying transition from laminar to turbulent flow. It is known that eddy-viscosity-based turbulence models cannot capture laminar-turbulence transition accurately. For this particular set-up, the URANS and hybrid LES/RANS model result with a zero modeled energy reducing the simulation to a very course DNS. Since a proper DNS is prohibitively expensive, and the RANS approach requires complicated stochastic models for particles to "feel" turbulence, LES stands as a common ground for simulating the flow inside 3D cavities. However, to properly resolve the boundary layer, LES can have stringent resolution requirements which scale with Reynolds number.

As an alternative, the hybrid RANS-LES models have been used in several studies to compute the flow in DHC but with the infinitely long cavity in the spanwise direction (y-direction), which helps sustain the modeled energy. The RANS-LES approach activates RANS mode near the wall – tolerating much coarser mesh near the wall – while LES is deployed when cell size is sufficient to resolve local motion scales in the far-wall region. In this light, [Abramov & Smirnov, 2006] studied the square cavity at a Rayleigh number of 1.58·10⁹ using Detached Eddy Simulation (DES) based on a one-equation model for the turbulent kinetic energy (TKE) transport. It was found in this study that although the mean flow was reasonably represented, turbulent statistics were observed to be poorly predicted in the near-wall proximity.

Recently Ali A.E.A et al., (2021) proposed a dual-mesh hybrid RANS-LES approach that can predict the flow to good accuracy in DHC, again with the periodic boundary condition in the y-direction. The idea behind that approach is based on using two overlapping computational domains, one for LES mode and the other is used to solve RANS equations. The main motivation for this study was to avoid the mismatch that often occurs at the LES/RANS interface when a single mesh is used for both models. It was reported that this approach yields satisfactory results in comparison with standard LES or RANS tested at the same study. However, such a method may raise many flags for future dispersed phase analyses that favor single-mesh simulation and consistent flow time and length scales.

We applied two widely used subgrid-scale models but on the coarse mesh. In the present work, this approach is investigated through the assessment of both the standard and dynamic Smagorinsky SGS models. This study aims to test the subgrid-scale model accuracy for the mesh resolution, which is below generally accepted LES standards.

The flow inside the cavity is buoyancy-driven where the vertical walls are held at two different temperatures and the rest of the walls are assumed to be insulated. As reported in the literature, it is very crucial to account for wall-to-wall radiation (between the bottom and top walls) [J. Kalilainen et al., (2016); Sergent A. et al., (2013)]. We therefore implicitly account for such effect by imposing the measured temperature profiles from the experiment as Dirichlet boundary conditions as suggested by Dehbi et. al., (2017). In a quantitative fashion, we report flow first and second-order statistics i.e. mean and RMS velocity profiles as well as mean temperature profiles. Results are compared against reference LES predictions by Dehbi et. al., (2017) and experimental measurements by J. Kalilainen et al, (2016). In particular, the high-quality particle image velocimetry (PIV) experimental measurements produced by J. Kalilainen made it possible to quantitatively validate our results.

In addition, we qualitatively show the main characteristics of the flow circulation at the cavity corners – where the flow is highly turbulent, as well as other flow structures.

## 2 Methodology

To compute the flow field we consider the momentum, continuity, and energy conservation equations to solve the incompressible non-Newtonian flow. Density variations due to thermal stratification are accounted for using the Boussinesq approximation for buoyancy. In this study, we use T-Flows [Niceno, B. et al, (2001)], which is a second-order accurate, unstructured, cell-centered, finite volume, in-house CFD code.

$$\frac{\partial u_i}{\partial t} + u_j \frac{\partial u_i}{\partial x_j} = -\frac{1}{\rho}\frac{\partial p}{\partial x_i} + \frac{v \partial^2 u_i}{\partial x_i \partial x_j} + \beta g_i(T - T_{ref}) \quad (1)$$

$$\frac{\partial u_i}{\partial x_i} = 0 \quad (2)$$

$$\frac{\partial T}{\partial t} + u_j \frac{\partial T}{\partial x_j} = \frac{\partial}{\partial x_i}\left(\frac{v}{Pr}\frac{\partial T}{\partial x_i} - \underline{\theta u_i}\right) \quad (3)$$

We use the Linear Eddy Viscosity (LEV) approximation to close the Reynolds stress term that stems from averaging the non-linear advection term in the momentum conservation equation Eq. (1). Using the Boussinesq eddy viscosity assumption, the stress tensor is evaluated by the space-filtered velocity field, which is linked to the mean velocity gradients. Using this hypothesis, the Reynolds stress tensor reads:

$$\tau_{ij} = -\underline{u_i u_j} = v_t\left(\frac{\partial u_i}{\partial x_j} + \frac{\partial u_j}{\partial x_i}\right) - \frac{2}{3}k\delta_{ij} \quad (4)$$

where k is the turbulent kinetic energy (TKE) and $\delta_{ij}$ is the Kronecker delta. The subgrid-scale (SGS) eddy viscosity $v_t$ is then computed based the filter width and the Smagorinsky parameter, which is computed in the standard model as follows

$$v_{t,LES-classic} = (\Delta_{LES} C_s)^2 |S| \quad (5)$$

where $C_s$ is the Smagorinsky constant which takes a fixed value between 0.1 - 0.2, while the Smagorinsky dynamic model allows the Smagorinsky constant to vary in space and time depending on an algebraic identity between the subgrid-scale stresses at two different filtered levels and the resolved turbulent stresses [Germano M. et al., (1991)].

$$v_{t,LES-Dynamic} = (\Delta_{LES})^2 C_{dyn} |S| \quad (6)$$

For energy transport (Eq. 3), we model the turbulent heat flux (THF) through the Simple gradient diffusion hypothesis (SGDH) as follows

Table 1: Fluid parameters and reference values

| Variable | $\rho$ | $\lambda$ | $\beta$ | $\upsilon$ | $c_p$ | $T_{ref}$ | $\rho_{ref}$ |
|---|---|---|---|---|---|---|---|
| Values | 1.0 | 2.434E-5 | 0.00319 | 1.72814E-5 | 1.0 | 39.18 | 1.0 |

$$u_i \theta = -\frac{\nu_t}{Pr_t}\frac{\partial T}{\partial x_i} \tag{7}$$

In the above equation, $Pr_t$ is the turbulent Prandtl number and it is fixed to the standard value of 0.9 in all simulations, while $C_{dyn}$ is the dynamic Smagorinsky parameter, and $\Delta_{LES}$ is the standard definition of LES cut-off length (i.e. the cubic root of the volumetric cell), and $|S| = \sqrt{2S_{ij}S_{ij}}$ is the magnitude of the strain rate, $S_{ij}$. It will be shown later in Section 4 that the Smagorinsky dynamic model can predict the correct eddy viscosity needed to damp fluctuations in the near-wall region, and hence, give a better prediction for higher moment central statistics in the near-wall region.

We consider a cubical cavity (x: horizontal, y: depth, z: vertical) with side length H = 0.7m, where all walls are set to no-slip boundary condition. The fluid flow is assumed homogeneous, incompressible, and Newtonian with constant viscosity and thermal diffusivity. Regarding thermal boundary conditions, vertical opposite walls are held at constant temperatures with $\Delta T=39.18$ to have a Rayleigh number of Ra= $10^9$. Like the LES study by Dehbi et. al., (2017), the front and back walls were set to adiabatic (passive) walls.

As shown in Fig. 1, the left vertical wall (x = 0) is assigned to be the hot wall (at temperature T = 330.54K), while the cold wall is at x=0.7 (with T = 291.36K). It was mentioned by J. Kalilainen et al., (2016) that the isothermal walls in the DIANA experiment were carried out within 0.4K uncertainty margin. Also, due care was taken to reduce heat losses from passive walls to the surroundings for which limit, the adiabatic wall assumption can still hold. To implicitly emulate wall-to-wall radiation effects, we set the temperature of bottom and top walls independently of the spanwise direction (y-axis). This is done by imposing the measured temperature profiles (from the DIANA experiment) as Dirichlet boundary conditions on the horizontal walls. This set of boundary conditions are called Intermediate Realistic conditions (IRC) as first coined by Xin et al., (2013). In that study, the authors showed that IRC can represent the right turbulent fields in the DHC.

As indicated in Table 1, the fluid parameters used for all simulations are fixed. For pressure momentum coupling, we use SIMPLE scheme and for time integration, we use the parabolic scheme. The central differencing scheme was used for the momentum equation while SMART scheme was used for the energy equation.

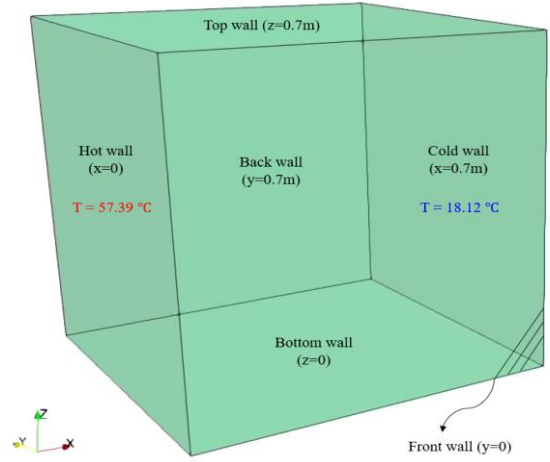

Figure 1: Schematic showing the geometry of the cavity with the thermal boundary conditions indicated.

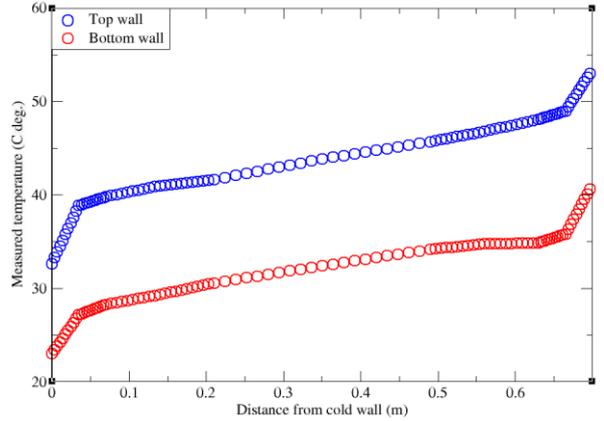

Figure 2: Measured temperature profiles for bottom and top walls from exp. Kalilainen et al., (2016).

For near-wall grid spacing, we use 0.003643 H for the first cell size. As shown in the Table 2 below, the corresponding y+ values for both SGS models are shown. The same mesh was used for each model with 56 nodes in each direction (the total number of cells is 166,375 cells). To guarantee a time-accurate solution, the time step size was assigned in each case so that the Courant–Friedrichs–Lewy number (CFL) is less than one throughout the simulation in all reported cases. It can be noticed from the table below that the dynamic Smagorinsky model takes about 33% extra CPU time compared to the standard Smagorinsky model. This is mainly due to the additional loops designed for the dynamic calculation of the model coefficient.

Table 2: CPU hours needed for VLES simulations

| Turbulence model | $Y_1^+$ | CPU [hr:min:sec] |
|---|---|---|
| Smagorinsky classic | 1.5 | 07:31:59 |
| Smagorinsky dynamic | 1.7 | 10:08:00 |

## 3 Flow field results

In this section, we show the results from both the standard Smagorinsky and the Dynamic models. It is usually standard to judge the predictions of the flow statistics after a certain number of flow time units, τ. In this flow configuration, the time unit is defined in terms of cavity length and circulation speed as follows:

$$V_r = \frac{\alpha \sqrt{Ra}}{L} \qquad (8)$$

$$\tau = \frac{L}{V_r} \qquad (9)$$

In the LES study of [Puragliesi et al., (2010)], authors consider 450 time units to assume statistically stationarity (fully developed flow), while both studies by Dehbi et. al., (2017) and Sergent A. et al., (2013) quote 600 time units. In this study, we adopt double the latter timespan, where turbulent statistics were engaged after 1200 time units from the beginning of the simulation, then results were averaged over another 1200 time units. This time-averaging interval was found to ensure full convergence of the flow. For our simulations, the circulation velocity and the time unit have values of 1.1 m/s and 0.637s respectively. Considering the time step size of 0.01s, this translates to approximately 153,000 time steps to cover the whole simulation from scratch until convergence.

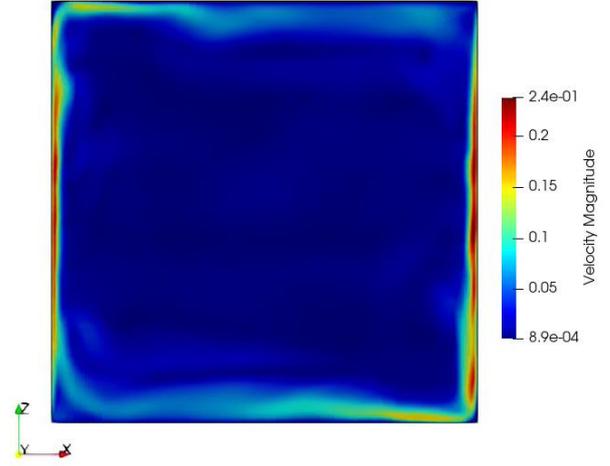

Figure 4: Contours of instantaneous velocity magnitude at the cavity midplane (x=0.35m) by the Smagorinsky dynamic model after 2400τ

The flow configuration inside DHC has a unique nature. This can be noticed from both the temperature and velocity contours shown in Fig. 3-4 where the highest velocity is located near the active cold and hot walls, while in the bulk region the fluid velocity almost nullifies. This can also be observed from the velocity contours shown in Fig. 5-6. The reason for this particular flow organization is due to the buoyancy force emerging from thermal stratification. It can also be noted that velocity gradients are quite high at the cavity corners due to the strong flow re-circulations as seen from the coherent turbulent structures in Fig. 7. This unique flow configuration makes the choice of the sub-grid scale model for wall treatment quite crucial. For this sake, we study with scrutiny the prediction of both the standard and dynamic Smagorinsky models to obtain credible databases.

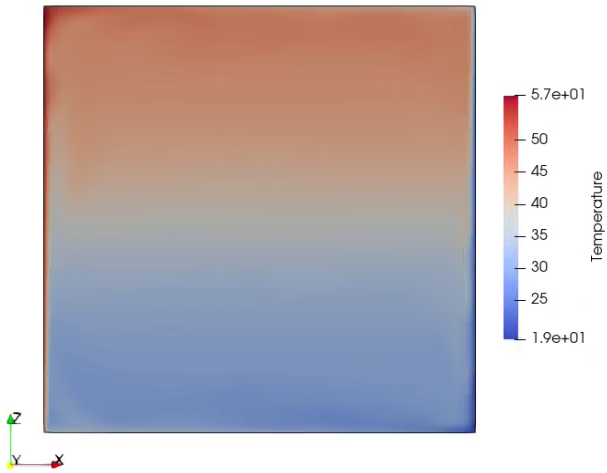

Figure 3 Contours of mean temperature at the cavity midplane (x=0.35m) by the Smagorinsky dynamic model after 2400τ

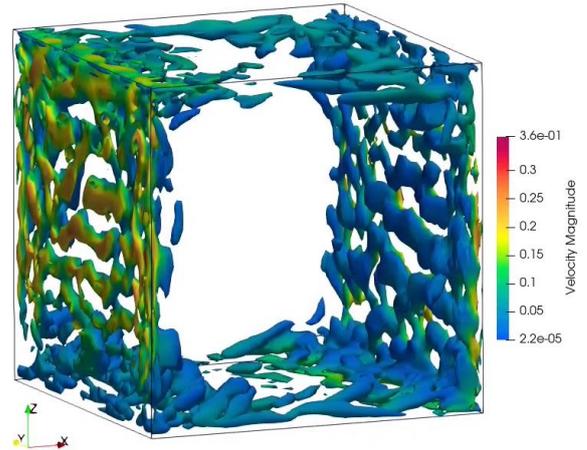

Figure 5: Plot of iso-surfaces of the normalized Q-criterion for a value of Qn = 0.65. The vortices are colored by velocity magnitude. Turbulent structures are produced by the standard Smagorinsky model with Cs=0.1.

As shown below, the mean and root mean square (RMS) profiles are plotted for both models to reveal the effect of the SGS model on flow prediction in the near-wall region – which is our main interest for the long run in terms of particulate cavity investigation. The mean temperature profiles give a very good match with the reference data, see Fig. 6. It can also be observed from Fig. 7 that mean vertical velocity profiles predicted by both SGS models are closely matching both LES and experimental data. However, the dynamic model gives a more accurate representation of the horizontal velocity near the wall.

For the higher moment statistics i.e. RMS values, the SGS model effect is more pronounced. It was shown that the vertical velocity fluctuation (RMS of w') is better captured by the dynamic model than the standard Smagorinsky model. This can be seen in Fig. 9 where the dynamic model captures 93% of the peak value of the velocity fluctuations, while the classic model predicts only about 67% of the reference value near the wall.

As mentioned above, the main reason why the dynamic Smagorinsky model performs better than the standard one is that the dynamic Smagorinsky parameter allows a better prediction for turbulent eddy viscosity values near the wall – and therefore offers a better SGS model for the near-wall region. This was also confirmed by calculating the eddy-over-molecular-viscosity ratio the value produced by the dynamic model was approximately 4 times higher than the one predicted by the standard model. This in turn damps turbulent fluctuations near the wall in the case of the dynamic model, and hence better represent the flow behavior.

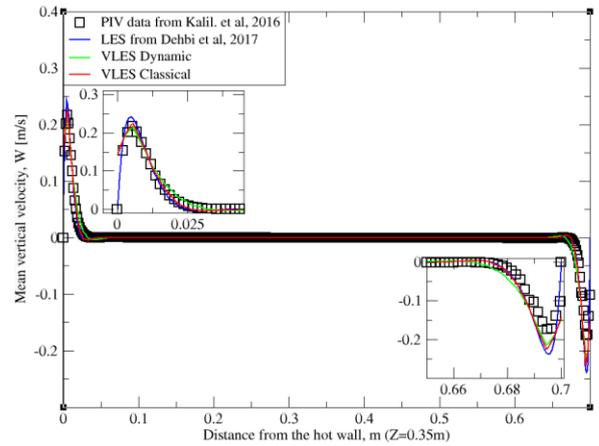

Figure 7: Comparison for mean vertical velocity profiles between hot and cold walls from both Smagorinsky models against reference LES and experimental databases. Profiles are obtained at (z=0.35m) using same mesh after 2400 time units.

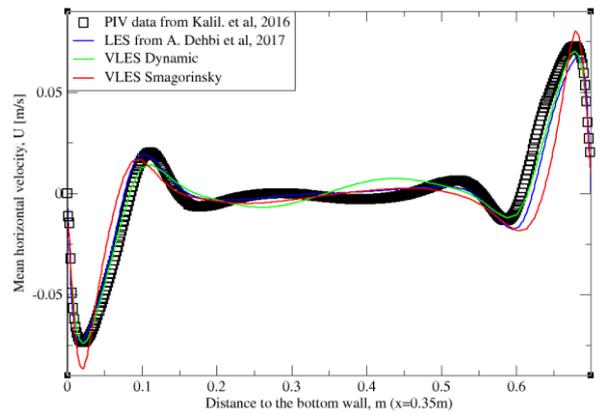

Figure 8: Comparison for mean horizontal profiles between bottom and top walls from both Smagorinsky models against reference LES and experimental databases. Profiles are obtained at (z=0.35m) using same mesh after 2400 time units.

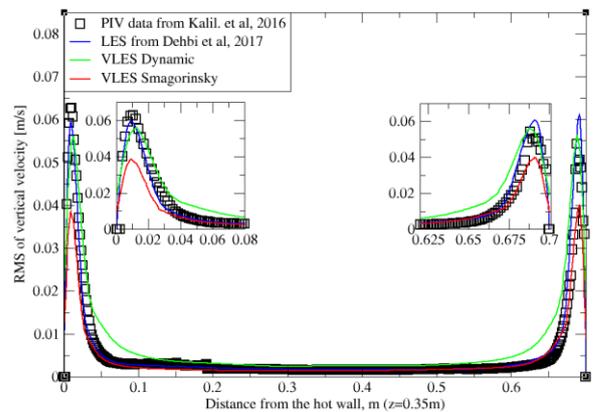

Figure 9: Comparison for RMS vertical velocity between hot and cold walls from both Smagorinsky models against reference LES and experimental databases. Profiles are obtained at (z=0.35m) using same mesh after 2400 time units.

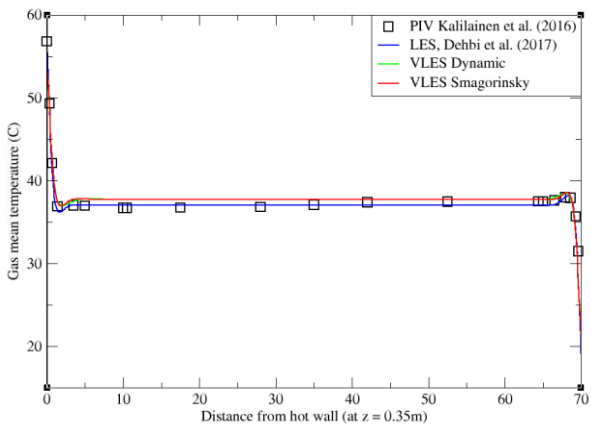

Figure 6: Comparison for mean temperature profiles between hot and cold walls from both Smagorinsky models against reference LES and experimental databases. Profiles are obtained at (z=0.35m) using same mesh after 2400 time units.

## Conclusions

The flow field inside a three-dimensional differentially heated cavity (DHC) has been investigated using large eddy simulations on a coarse mesh at Rayleigh number of $Ra = 10^9$. In particular, two LES models have been used namely: the standard Smagorinsky and dynamic Smagorinsky models. Wall-to-wall radiation effects have been implicitly taken into account by imposing the PIV-measured temperature profiles of bottom and top walls as Dirichlet BCs in all simulations. A quantitative analysis is reported for the fluid flow where first and second-moment statistics are analyzed. Results are compared against both well-resolved LES by Dehbi A. et. al., (2017) and the experimental database by Kalilainen J. et al, (2016).

It was shown that the models could predict the mean flow properly even when a coarse mesh is used. However, for high moments i.e. RMS values for horizontal and vertical velocity fluctuations, a significant underprediction of the flow statistics were observed in the near-wall region by the standard Smagorinsky model. On the other hand, the Dynamic model predictions match the reference LES and experimental databases very well. The main reason behind this difference is that the eddy viscosity values are better predicted near the wall by the dynamically adjusted model's coefficient, and hence the flow is well represented compared to the standard model. First and second central moment statistics are globally in a very good match with reference LES and experimental measurements at a fraction of CPU cost relative to LES by A. Dehbi et al., (2017). Good performance of coarse LES simulation with less than 0.2 million compared to the fine 5.1 million-cell-mesh of well-resolved LES confirms that for the cavity simulation LES with the right SGS model still can be used which is a promising step for further multiphase flow investigations.

## Acknowledgements

This research project is sponsored by Swiss National Science Foundation (SNSF) under grant number [200021175532] in membership with the Federal pension fund PUBLICA according to the regulation of ETH domain (VR-ETH 1).

## References

Abramov AG, Smirnov EM. Numerical simulation of turbulent convection of air in a square cavity heated on the side. High Temp 2006;44(1):91–8.

Ali A.E.A., Afgan I, Laurence D, Revell A, A dual-mesh hybrid RANS-LES simulation of the buoyant flow in a differentially heated square cavity with an improved resolution criterion, Computers and Fluids 224 (2021) 104 94 9.

Dehbi, A., Kalilainen, J., Lind, T., Auvinen, A., A large eddy simulation of turbulent particle-laden flow inside a cubical differentially heated cavity, J. Aero. Sci., 103 (2017), pp. 67-82.

De Vahl Davis, G. Laminar natural convection in an enclosed rectangular cavity, International Journal of Heat and Mass Transfer 11 (1968) 1675–1693.

Germano, M., Piomelli, U., Moin, P. and Cabot, W. H. (1991), "A Dynamic Subgrid-Scale Eddy Viscosity Model", Ph. of Fluids A, Vol. 3, No. 7, pp. 1760-1765.

Jones, A. V., & Kissane, M. (2000). State of understanding of fission product transport in the circuit and of aerosol behaviour in the containment of Phebus. Journal of Aerosol Science, 31(Supp.1), S37–S38.

Kalilainen, J., P. Rantanen, T. Lind, A. Auvinen, A. Dehbi, Experimental investigation of a turbulent particle-laden flow inside a differentially heated cavity, J. of Aero Sci., 100 (2016), pp. 73-87.

Kam, T., Hsia, L. C., & Chang, T. (1998). Clean room particle monitor, airflow simulation and measurement for aerosol reduction. J. of Aerosol Science, 29, 254.

Morrison, G. C., Zhao, P., & Kasthuri, L. (2006). The spatial distribution of pollutant transport to and from indoor surfaces. Atmospheric Environment, 40, 3677–3685.

Niceno, B., Hanjalic, K. e.a., 2001. T-flows. https://github.com/DelNov/T-Flows.git/.

Poussou, S. B., Mazumdar, S., Plesniak, M. W., Sojka, P. E., & Chen, Q. (2010). Flow and contaminant transport in an airliner cabin induced by a moving body: Model experiments and CFD predictions. Atmospheric Environment, 44, 2830–2839.

Puragliesi, R. (2010). Numerical Investigation of Particle-Laden Thermally Driven Turbulent Flows in Enclosure, Ph.D. thesis, Eo: 4600. Ecole Polytechnique Federale de Lausanne, Lausanne.

Sebilleau, F., Issaa R, Lardeau S, Walker S. Direct Numerical Simulation of an air-filled differentially heated square cavity with Rayleigh numbers up to 1E11. Int J Heat Mass Transf 2018; 123:297-319.

Sergent, A., Xin, S., Joubert, P., Le Quéré, P., Salat, J., & Penot, F. (2013). Resolving the stratification discrepancy of turbulent natural convection in differentially heated air filled cavities – Part I: Reference solution using Chebyshev spectral methods. International Journal of Heat Fluid Flow, 39, 1–14.

Wana, M. P., Chaoa, C. Y. H., Nga, Y. D., Sze Toa, G. N., & Yub, W. C. (2007). Dispersion of expiratory droplets in a general hospital ward with ceiling mixing type mechanical ventilation system. Aerosol Science and Technology, 41, 244–258.

Xin, S., Salat, J., Joubert, P., Sergent, A., Penot, F., & Le Quéré, P. (2013). Resolving the stratification discrepancy of turbulent natural convection in differentially heated air-filled cavities. Part III: A full convection-conduction-surface radiation coupling. International Journal of Heat Fluid Flow, 42, 33–48.